\documentclass[fleqn,usenatbib]{mnras}

\usepackage{newtxtext,newtxmath}

\usepackage[T1]{fontenc}
\usepackage{ae,aecompl}

\usepackage{graphicx}	
\usepackage{amsmath}	
\usepackage{amssymb}	

\newcommand{\deriv}[2]{\frac{{\rm d} #1}{{\rm d} #2}}
\newcommand{\pderiv}[2]{\frac{\partial #1}{\partial #2}}

\title[Fingerprints of giants]{Fingerprints of giant planets in the composition of solar twins}

\author[Booth \& Owen]{
Richard A. Booth$^{1}$\thanks{E-mail: rab200@ast.cam.ac.uk (RAB)}
and  James E. Owen$^{2}$
\\
$^{1}$Institute of Astronomy, University of Cambridge, Madingley Road, Cambridge, CB3 0HA, UK\\
$^{2}$Astrophysics Group, Imperial College London, Blackett Laboratory, Prince Consort Road, London SW7 2AZ, UK
}

\date{Accepted 2020 February 24. Received 2020 February 14; in original form 2019 December 20}

\pubyear{2020}

\begin{document}
\label{firstpage}
\pagerange{\pageref{firstpage}--\pageref{lastpage}}
\maketitle

\begin{abstract}
The Sun shows a $\sim 10$~\% depletion in refractory elements relative to nearby solar twins. It has been suggested that this depletion is a signpost of planet formation. The exoplanet statistics are now good enough to show that the origin of this depletion does not arise from the sequestration of refractory material inside the planets themselves. This conclusion arises because most sun-like stars host close-in planetary systems that are on average more massive than the Sun's. Using evolutionary models for the protoplanetary discs that surrounded the young Sun and solar twins we demonstrate that the origin of the depletion likely arises due to the trapping of dust exterior to the orbit of a forming giant planet. In this scenario a forming giant planet opens a gap in the gas disc, creating a pressure trap. If the planet forms early enough, while the disc is still massive, the planet can trap $\gtrsim 100$~M$_\oplus$ of dust exterior to its orbit, preventing the dust from accreting onto the star in contrast to the gas. Forming giant planets can create refractory depletions of $\sim 5-15\%$, with the larger values occurring for initial conditions that favour giant planet formation (e.g. more massive discs, that live longer). The incidence of solar-twins that show refractory depletion matches both the occurrence of giant planets discovered in exoplanet surveys and ``transition'' discs that show similar depletion patterns in the material that is accreting onto the star.
\end{abstract}




\section{Introduction}

High resolution spectroscopy has enabled precise differential measurements of stellar composition, which led to the discovery that the Sun's surface composition is anomalous when compared to the solar twins (stars with almost identical surface temperature, gravity, and [Fe/H] to those of the Sun; \citealt{Melendez2009}). Compared with 80 to 90 per cent of solar twins, the abundances of refractory (rock-forming) species in the Sun are depleted with respect to iron and more volatile elements (such as carbon and oxygen), with the depletion showing a correlation with condensation temperature of each element \citep{Melendez2009, Ramirez2009, Gonzalez2010, GonzalezHernandez2010, GonzalezHernandez2013, Schuler2011b, Bedell2018}. This result sparked numerous potential explanations for the condensation temperature trend, including the depletion due to the formation of rocky planets \citep{Melendez2009} or enrichment by planet engulfment \citep{Ramirez2011, Spina2015, Oh2018, Church2020}. Alternative ideas include the removal of dust from the parent molecular cloud \citep{Gustafsson2010} or variations induced by Galactic chemical evolution \citep{Adibekyan2014, Nissen2015, Spina2016}. However, recently \citet{Bedell2018} showed that the condensation temperature trend is still present after correcting for Galactic chemical evolution.

While depletion by locking up refractory material in planets may seem attractive, there are several difficulties associated. The motivation for the planet formation idea is that the depletion pattern discovered by \citet{Melendez2009} is well reproduced by a composition made up of equal parts Earth-like and chondritic composition \citep{Chambers2010}. Although \citet{Chambers2010} showed that the 0.04 dex (10 per cent) depletion can be explained if $4\,M_\oplus$ of rocky material has been removed from the surface convective zone of the Sun, the convective zone of a young $1\,M_\odot$ star must be much larger than the Sun's 2.5 per cent by mass to explain the luminosity spread in young stellar clusters \citep{Baraffe2012}. Although there is still some uncertainty, recent results suggest that convective zone was likely 50 to 100 per cent by mass \citep{Kunitomo2018} during the time while the Sun was surrounded by a protoplanetary disc (e.g. ages $\lesssim 10$~Myr; \citealt{Haisch2001},  although note that disc lifetimes are environment dependent \citealt{Pfalzner2014,Winter2018}), resulting in the need to produce differences of up to $160\,M_\oplus$ of rocky material (i.e. up to 10 per cent of the rocky material; \citealt{Lodders2003}). This is certainly more than the mass in the rocky planets and asteroids of the solar system. While the gas giants and ice giants may contain the necessary mass in solids, their composition does not naturally produce the condensation temperature trend because they likely contain a substantial component of carbon and oxygen containing volatile ices (see \citealt{Kunitomo2018}, for a discussion). Finally, (as we will explicitly demonstrate in this work) planet formation is now known to be common. The fraction of Sun-like stars that contain planets is certainly $\gtrsim 50\%$ \citep{Fressin2013} and could range between 70\% to 100\% \citep{Mulders2018,Zink2019} depending on the inclination distribution of multi-planet systems. Therefore, the vast majority of the solar-twins probably host super-Earth/mini-Neptune planetary systems that are likely much more massive than our terrestrial planetary system  \citep[e.g.][]{Chiang2013}. This means that low mass, rocky planet formation cannot be the origin of the abundance anomaly.  Instead, an alternative explanation seems more likely.

We explore whether the depletion of refractory elements can be explained via the segregation of dust and gas in protoplanetary discs (as suggested by \citealt{Gaidos2015}) associated with the formation of giant planets. In this scenario the giant planets form early, while the disc is still relatively massive. These planets open gaps in the protoplanetary disc \citep{Lin1986,Crida2006}, trapping dust outside of their orbits in pressure traps \citep[e.g.][]{Rice2006,Pinilla2012,Zhu2012,Zhu2014}, while allowing gas to flow through. By preventing dust from reaching the star, giant planets will cause the star to be depleted in refractory species. The gap opening by a giant planet scenario has several merits. First, giant planets are considerably rarer than super-Earths/mini-Neptunes \citep[e.g.][]{Fernandes2019} with the frequency of Jupiter analogues being around 5-10\% \citep{Wittenmyer2016,Wittenmyer2019}, similar to the fraction of anomalously depleted solar twins. Second, since the dust traps created by giant planets are quite efficient, they can (depending on the timescale for planet formation) prevent a large fraction of the disc's initial refractory reservoir accreting on to the star. Finally, there is observational evidence of refractory material being trapped in the outer regions and prevented from accreting on the star occurring in observed protoplanetary discs.

Observations of protoplanetary discs show that $\sim 10-20$ per cent have large gaps or cavities \citep[e.g.][]{Kenyon1995,Koepferl2013,vanderMarel2018}. These discs are known as ``transition discs''. The origin of transition discs is still debated \citep[e.g.][]{Owen2016,Ercolano2017b} and it is clear they are not a homogeneous class of object. However, the large radii ($\gtrsim 10$~AU) of some of these cavities, and the ongoing gas accretion onto the star points to planets as their origin \citep[e.g.][]{OwenClarke2012}. These, ``large-holed'' transition discs typically have some of the highest dust masses of any discs ($\gtrsim 100$~M$_\oplus$ of solids,  e.g. \citealt{Andrews2012}), placing them among the most favourable objects given the large depletion required. Third, transition discs likely produce the required condensation temperature trend, as the refractories remain trapped, yet volatiles can enter the gas phase and flow through the planet gap onto the star. A key piece of evidence in favour of this hypothesis comes from transition discs around young stars more massive stars than the Sun. Since young A stars (Herbig Ae stars) have only a thin convective zone at their surfaces, their surface composition provides a record of their recent accretion history \citep[e.g.][]{Jermyn2018}. \citet{Kama2015} showed that the material Herbig Ae stars accreted from  transition discs with large cavities was typically depleted in refractory elements by about a factor of 10 (but not carbon or oxygen), while stars accreting from normal, non-transition discs showed a solar composition. This was true even for discs with cavities of several $10\,{\rm AU}$. It is much more challenging to conduct the same experiment for T Tauri stars, given that their convective envelopes erase the record of their current accretion, but the line ratios of emission coming from the material in their accretion columns suggests that a similar depletion also occurs \citep{Booth2018}. Together these factors make a compelling case for transition discs, associated with the formation of giant planets as the origin of refractory depletion.

In this paper, we model the evolution of gas and dust in protoplanetary discs, exploring how the presence of planets affects the composition of the stellar photosphere relative to cases that do not form planets. Firstly, we demonstrate explicitly that the formation of terrestrial rocky planets cannot be the origin of the depletion when taking into account the exoplanet statistics from {\it Kepler}. Secondly, we show that trapping of dust outside the orbit of a forming giant planet can reproduce the required refractory depletion provided the planet forms early enough. Section \ref{sec:model} describes our model in detail, while Section \ref{sec:results} shows that this scenario comfortably reproduces the required depletion of refractory elements rather than sequestration of refractory material inside planets. Finally in sections \ref{sec:discuss} and \ref{sec:conclusion} we discuss the implications of our results and present our conclusions.

\section{Model}
\label{sec:model}

Our model for the depletion of refractory elements is based upon two components: a model for the evolution of gas and dust in the protoplanetary disc and a model for the evolution of the solar convective zone. The first model provides the amount of dust and gas accreted onto the star over time, while the second model converts this accretion into a predicted stellar abundance. We compare models with and without a giant planet to determine the influence of giant planets on the composition of the solar twins. In this work we make the crude assumption that all the refractory elements and contained in the dust and all the volatile elements are contained in the gas. This simplification is necessitated by the limited understanding of chemical transport in protoplanetary discs, especially when considering the interaction with forming planets. Therefore, we do not attempt to model the trend with condensation temperature (see Section~\ref{sec:discuss} for speculations along these lines), rather we just consider the total ``refractory'' depletion, via the total dust depletion.

We model the evolution of the protoplanetary disc assuming the standard paradigm for disc evolution. Namely the gas evolves viscously, and disc dispersal occurs via of photoevaporation \citep[e.g.][]{Armitage2011}. This standard model is known to reproduce the protoplanetary disc evolution diagnostics around T Tauri stars such as the disc lifetimes, disc fractions, gas accretion rates and loci discs transverse in infra-red colours, see a recent review by \citet{Ercolano2017b}. Almost all of our base model parameters are taken from the calculations of \citet{Owen2011} which calibrated the initial conditions and evolutionary parameters to the disc evolution diagnostics. Our dust disc model includes radial drift, diffusion and growth/fragmentation and co-evolves with the gas disc. The initial gas and dust disc surface densities are chosen to follow a zero-time Lynden-Bell \& Pringle similarity solution \citep{LyndenBell1974}. This class of models has been detailed in the literature many times, thus the busy reader may skip this section; however, we discuss it below because this the first time that all of these effects have been included together.

As discussed in the introduction we create a transition disc through gap-opening by a forming giant planet. In our model, gap opening by the planet is treated by including the influence of the torques from the planet on the disc, as in \citet{Alexander2009}. The surface density of the gas, $\Sigma_g$, and dust $\Sigma_d$, evolve according to 
\begin{equation}
\pderiv{\Sigma_g}{t} = \frac{1}{R} \pderiv{}{R}\left[3 R^{1/2}\pderiv{}{R}\left(\nu \Sigma_G R^{1/2}\right) - \frac{2 \Lambda \Sigma_g}{\Omega} \right] - \dot{\Sigma}_w,
\end{equation}
and 
\begin{equation}
\pderiv{\Sigma_d}{t} = - \frac{1}{R} \pderiv{}{R}\left[R v_d \Sigma_d - \frac{\nu}{Sc} R \Sigma_g \pderiv{}{R}\left(\frac{\Sigma_d}{\Sigma_g}\right) \right].
\end{equation}
Here $R$ is the radius, $t$ is the time, $\nu  = \alpha c_s H$ is the viscosity (where we take $\alpha = 2.5 \times 10^{-3}$), $\Omega$ is the Keplerian angular frequency, $\Lambda$ is the torque from a planet, and the disc temperature profile follows $T\propto R^{-1/2}$ with the temperature at 1~AU set to 100~K. We take the Schmidt number, $Sc$, to be unity. For the mass-loss rate due to photoevaporation, $\dot{\Sigma}_w$, we use the fits to the X-ray photoevaporation rates of \citet{Owen2010,Owen2011} given in \citet{Owen2012}, assuming an X-ray luminosities, $L_{\rm X}$, of $4.5\times10^{29}$, $1.5\times10^{30}$ and $3.25\times10^{30}$ erg~s$^{-1}$ which cover the T Tauri stars X-ray luminosity function \citep{Gudel2007}. These photoevaporation rates give disc lifetimes of roughly 10, 4, \& 2 Myr respectively.  The effect of photoevaporation on the dust surface density is neglected because dust settles to the disc mid-plane and only small dust particles can be carried away in the wind, thus the dust-mass lost in the wind is negligible \citep[e.g.][]{Owen2011b}. In the absence of a planet ($\Lambda=0$) the gas evolution model is nearly identical to \citet{Owen2011} and the gas and dust evolution model to \citet[][the calculations without radiation pressure]{Owen2019}, which reproduces the disc life-times, fractions and accretion rates in nearby star-forming regions. Since the \citet{Owen2011,Owen2019} calculations focused on a typical T Tauri star with a mass of 0.7~M$_\odot$, to achieve the same disc lifetimes we set the initial disc radius to be 45~AU. The planet the calculations are very similar to \citet{ErcolanoRosotti} who modelled the evolution of migrating, forming giant planets in the \citet{Owen2011} gas-only disc evolution model. 

The torque due to the planet, with a mass ratio of $q$, follows \citet{Owen2014}, using the symmetric form given by \citet{Armitage2002}:
\begin{equation}
\Lambda=
\begin{cases}
-\frac{q^2GM_*}{2R}\left(\frac{R}{\max(H,|R-a|)}\right)^4 & \text{if $R<a$},\\
\frac{q^2GM_*}{2R}\left(\frac{a}{\max(H,|R-a|)}\right)^4 & \text{if $R>a$}.
\end{cases}
\end{equation}
We allow the planet to migrate, computing its migration rate via conservation of angular momentum \citep[e.g.][]{Armitage2002}. Accretion onto the planet is included following \citet{Owen2014}, here we assume that 5\%  per cent of the gas flowing past the planet is accreted and we halt accretion once the mass of the planet reaches a Jupiter mass. These choices result in the planet reaching a Jupiter mass by 4-5 Myr and an orbital separation of 3-4~AU at the end of the simulations. We note the values of the dust and gas masses accreted by the star are highly insensitive to these choices. 

The radial velocity of the dust, $v_d$, is given by 
\begin{equation}
v_d =\frac{v_g St^{-1} - \eta \Omega R}{St + St^{-1}}, \label{eqn:d_vel}
\end{equation}
where $v_g$ is the radial velocity of the gas and 
\begin{equation}
\eta=-\frac{{\rm d}\log P}{{\rm d}\log R}\left(\frac{c_s}{v_K}\right)^2
\end{equation}
\citep{Takeuchi2002}. Here, $v_K = R \Omega$ is the Keplerian velocity and  $St$ is the Stokes number of the dust grains, evaluated in the disc mid-plane. Assuming the Epstein drag law and a Gaussian vertical structure for the disc, the Stokes number is given by
\begin{equation}
St = \frac{\upi}{2} \frac{\rho_s a}{\Sigma_g},
\end{equation}
where $\rho_s = 1.25\,{\rm g\,cm}^{-3}$ is the internal density of the grains and $a$ is their radius. We set the size of the dust grains according to the two-population model of \citet{Birnstiel2012}. We assume that the dust grains, with initial size $a = 0.1\micron$, grow on a time scale
\begin{equation}
    t_{\rm grow} = \frac{a}{\dot{a}} = f_{\rm grow} \left(\frac{\Sigma_d}{\Sigma_g} \Omega\right)^{-1}.
\end{equation}
Grain growth continues until it reaches one of two barriers: the `fragmentation barrier', or the `radial drift barrier'. Fragmentation occurs when turbulence drives collisions between dust grains to velocities that exceed a certain threshold speed, $v_f$, which limits the grains to a maximum Stokes number,
\begin{equation}
St_{\rm frag} = 0.37 \frac{1}{3\alpha}\frac{v_f^2}{c_s^2}, \label{Eqn:FragLimit}
\end{equation}
where the factor 0.37 comes from fits to detailed coagulation calculations \citet{Birnstiel2012}. We take $v_f = 10\,{\rm m\,s}^{-1}$, appropriate for an icy composition \citep[e.g.][]{Gundlach2015}. The radial drift limit occurs when the growth time exceeds the radial drift time of the dust grains, which occurs at the Stokes number
\begin{equation}
St_{\rm drift} = 0.55 \frac{1}{f_{\rm grow}} \frac{\Sigma_g}{\Sigma_d}  \left( \frac{R}{H} \right)^2 \left|\deriv{\ln P}{\ln R}\right|^{-1}, \label{Eqn:DriftLimit}
\end{equation}
with the factor 0.55 again from \citet{Birnstiel2012}. In our expressions for $t_{\rm grow}$ and $St_{\rm drift}$ we have included an extra factor $f_{\rm grow}$ to account for the known result that current grain growth models predict much shorter lifetimes for the dust in discs than the gas, while the observed lifetime of discs inferred from fraction of young stars with discs obtained from near infra-red excesses (which trace the dust) is similar to that still obtained from the fraction of stars with measurable accretion rates (which traces the gas) across regions of different ages \citep{Fedele2010}. Physically, $f_{\rm grow}$ can interpreted in terms of the fraction of collisions that lead to growth: $f_{\rm grow}=1$ means 100\% of collisions result in growth, whereas $f_{\rm grow}=100$ means only 1\% of collisions result in growth. We note the appropriate value of $f_{\rm grow}$ is highly uncertain and does quantitatively effect our results. 

\begin{figure}
    \centering
    \includegraphics[width=\columnwidth]{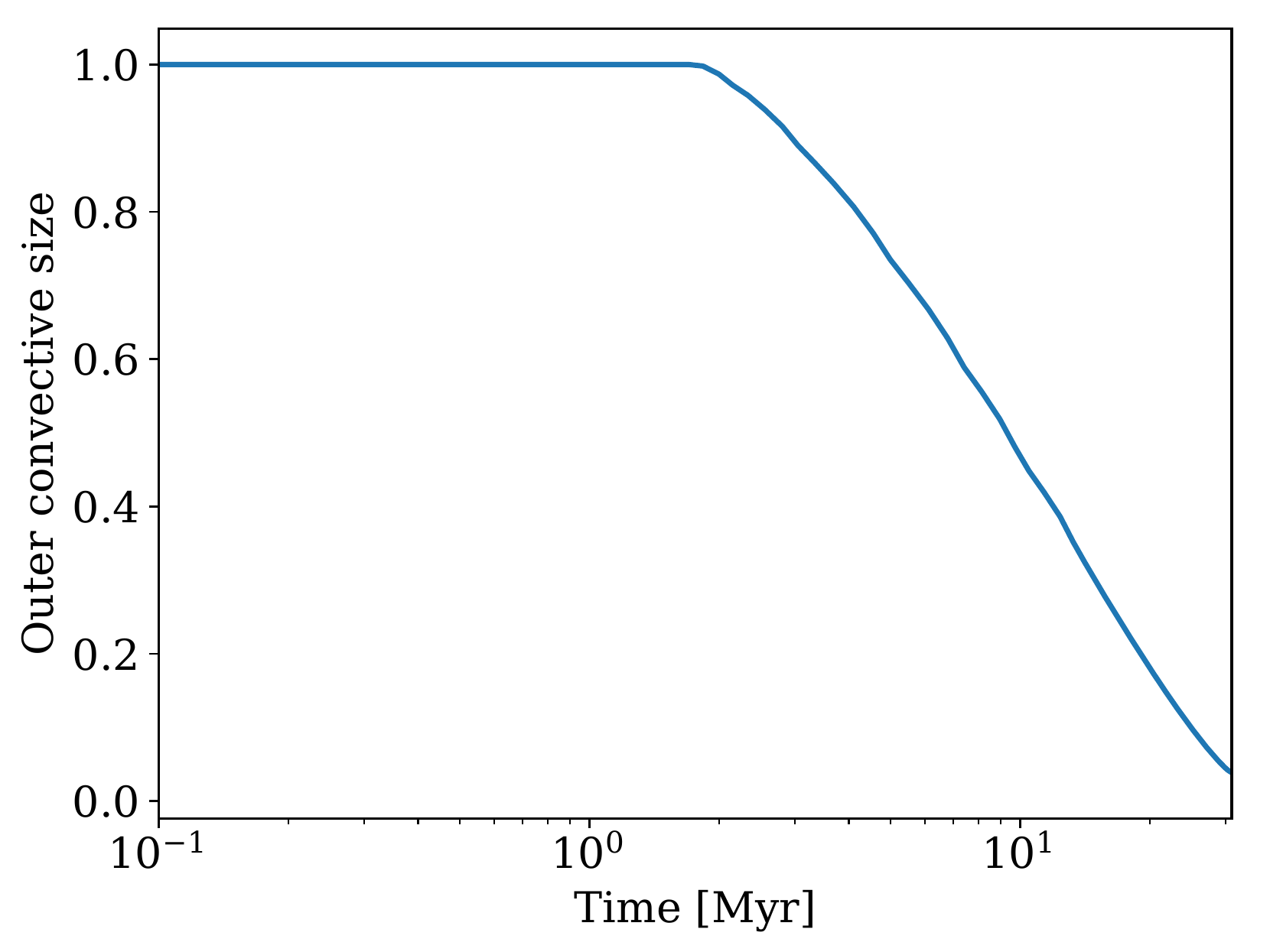}
    \caption{Fraction of stellar mass contained in the outer convective zone.}
    \label{fig:ConvZone}
\end{figure}

We begin our disc simulations at the end of Class 0/I phase, which we take to be a few $10^5\,{\rm yr}$, although our results are not sensitive to this value as long as it is $\ll 1\,{\rm Myr}$. We assume an initial disc mass of $0.1\,M_\odot$ of gas and the initial dust-to-gas ratio is taken to be 0.01, although we run one set of models with it set to 0.02. We explored planets forming between 15-25~AU, but present only the 20~AU case because the results are insensitive to this choice. The initial mass is taken to be 0.01~M$_J$. In most cases the planet forms 0.5~Myr after the simulation begins; however, we also model cases where it forms at the beginning of the simulation. It is critical in these calculation that the planet forms at large separations early. This is because a pressure trap is only efficient if the dust grains have $St \gtrsim \alpha$, which typically means their growth is drift-limited rather than fragmentation limited. Fragmentation limited dust is typically coupled to the gas and is thus not efficiently trapped. Dust is typically drift limited in the outer disc, due to the lower collision velocities between the grains \citep[e.g.][]{Birnstiel2012}.  

We run all of the simulations until accretion onto the star is terminated by photoevaporation, and hence no more gas or dust accretion can take place through the disc.

The abundance of the stellar photosphere is computed by tracking the composition of the stars outer convective zone.  The size of this convective region is computed using the MESA stellar evolution code \citep{Paxton2011,Paxton2013,Paxton2015} to evolve a $1\,M_\odot$ star, recording the size of its outer most convective zone (\autoref{fig:ConvZone}).  The evolution parameters are the default ones in MESA for the evolution of a Sun-like star. We assume that the disc and star initially have the same abundance. To track the abundance of the outer convective zone we assume that the mass accreted is immediately mixed throughout the convective zone. Once the star is no longer fully convective, we assume that the shrinking of the convective zone occurs without changes in composition. \citet{Kunitomo2018} showed that this procedure accurately reproduces the composition when compared to models that explicitly evolve the star during accretion. Our model for the surface composition neglects the influence of accretion on the evolution of the convective zone size, which has been shown to be sensitive to the accretion history \citep{Baraffe2010}. However, \citet{Kunitomo2017,Kunitomo2018} showed that models that reproduce the luminosity spread in young stellar clusters tend to have similar convective zone evolution to the non-accreting model used here.

\section{Results}
\label{sec:results}

\subsection{Depletion due to rocky planets}
\label{sec:rocky_planets}

\begin{figure*}
    \centering
    \includegraphics[width=\textwidth]{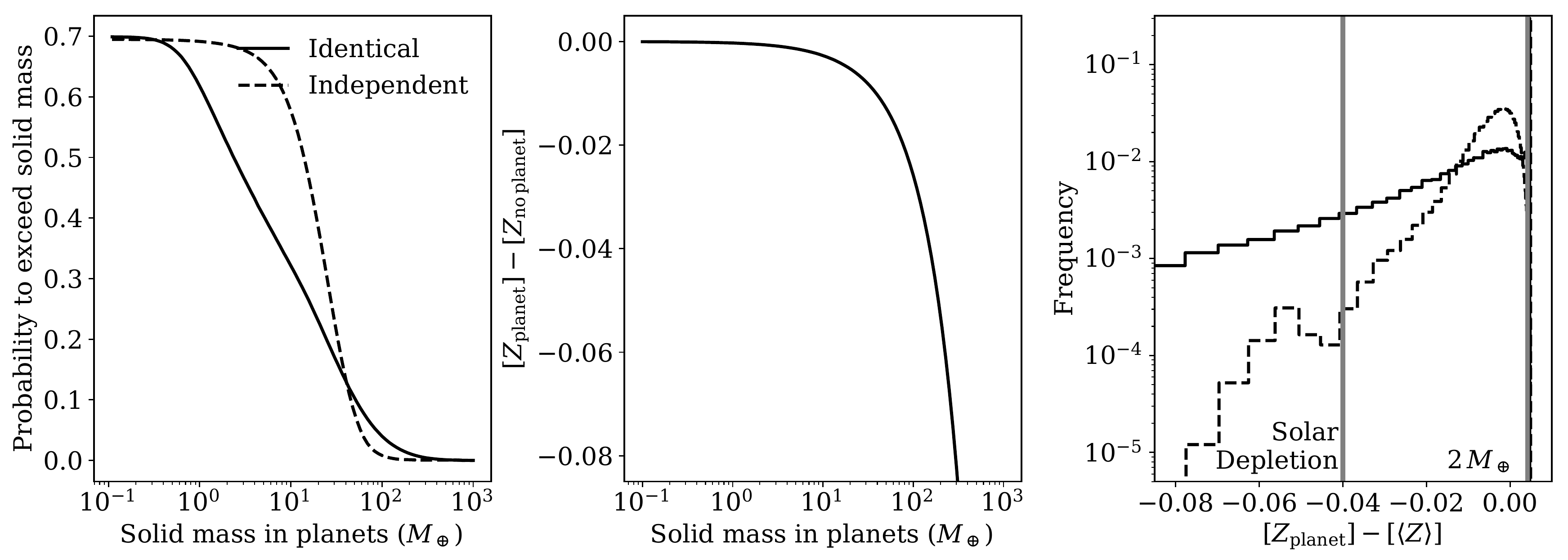}
    \caption{Effect of rocky planets on the photospheric abundance of refractory material. Left: Distribution of planetary system masses. Middle: photospheric depletion as a function of mass in rocky planets. Right: Distribution of photospheric depletions for systems that host planets relative the to average of all stars. The amount of solids locked up rocky planets in the solar system ($2\,M_\oplus$) and the depletion of the solar refractory abundances (10 per cent, 0.04 dex) are marked by vertical grey lines. }
    \label{fig:KeplerDist}
\end{figure*}

The original suggestion for the depletion was that the ``missing'' refractory material was locked-up in the Sun's rocky planets \citep{Melendez2009}. Exoplanet studies are now sufficiently complete to test  the hypothesis that the Sun and its twin stars formed from material with the same composition, with the differences in the refractory abundances today arises due to the material locked up into rocky planets.

To estimate the depletion caused by the mass locked up in typical rocky exoplanet systems we first make an estimate of the distribution of their masses. Since the exoplanet studies are dominated by transit surveys which only measure planetary radius, we do this by first generating synthetic planetary systems according to the radius and multiplicity distribution of \citet{Zink2019}. \citet{Zink2019} fit simultaneously for the radius distribution and multiplicity distribution of planets found by the {\it Kepler} satellite, accounting for the reduction in detection efficiency for each subsequently detected planet. If we used multiplicity distributions derived neglecting this effect we would recover lower system masses. Since the radii of planets within the same system are known to be mildly correlated \citep[e.g.][]{Weiss2018} we have tested the influence of correlated sizes by examining the two most extreme scenarios: where the sizes of each planet in a system are independent or identical. Finally, to convert the planet radii to masses we use \textsc{forecaster} \citep{Chen2017}, which we sum to generate the total mass in a given system.

The mass of depleted material is computed by optimistically assuming that the planets are made entirely from refractory material, i.e. we neglect the contribution of water that would reduce the depletion of refractory elements with respect to volatiles, although this is known to be small $\lesssim 10\%$ for the super-Earth/mini-Neptune population \citep[e.g.][]{Wolfgang2015}. To compute the photospheric depletion, we use the accretion history for our protoplanetary disc evolution model with median X-ray luminosity, no giant planets, and $f_{\rm grow}=10$, although the conclusions do not depend on the chosen model. The dust accreted is assumed to have a solar composition of refractory material \citep{Lodders2003} until a time $t_{\rm crit}$, after which the accretion of refractory material is assumed to be zero. The value of $t_{\rm crit}$ is chosen to give the required depletion. This choice of $t_{\rm crit}$ it the most optimistic possible choice for depletion possible as the convective zone it at its smallest when we prevent refractory material from being accreted. 

The results of this exercise are shown in \autoref{fig:KeplerDist}.  The left panel shows that most stars (50 to 70 per cent) should host more mass in rocky planetary systems than the Solar System, $\sim 2 M_\oplus$. In the middle panel we show the correspondence between the mass in the planetary system and the depletion compared to stars without planets, while the right panel shows the distribution of stellar depletions compared to the sample mean; in this space the Sun's depletion is 0.04 dex.  Since most planetary systems are expected to have more mass than the rocky Solar System bodies but lower depletions of refractory species in their photospheric abundances, it is clear that the mass of refractory material locked up in rocky bodies cannot be responsible for the observed depletion. Furthermore, the 0.01 to 0.02 dex depletion produced under the optimistic assumptions invoked here is small enough to not significantly affect any signal produced by gap opening due to giant planets. 

Although the estimates produced here argue against rocky planet formation as the origin of the refractory depletion, we briefly consider whether modifications to the above argument could be used to reconcile the differences. To change the conclusions, we would need to find a way to make Solar system amongst the most massive planetary systems, and increase to the photospheric depletion produced for a given mass of refractory material locked up into planets because only a few percent of Kepler systems produce a large enough depletion. If the sun had a smaller convective zone than we consider here, this would reduce the mass of refractory material needed, but constraints from the luminosity distribution in stellar clusters likely rule out convective zone sizes much smaller than we use here \citep[e.g.][]{Baraffe2012,Kunitomo2017},  meaning that at least several $\times 10\,M_\oplus$ of refractory material is likely required. Bringing the required depletion down to $30\,M_\oplus$ instead of $150\,M_\oplus$ would mean that approx 20 per cent of Kepler systems could produce the required depletion, but it would still not place the Solar system amongst the most depleted systems.

The Sun could be placed amongst the most depleted systems by considering the Solar System giant planets. The mass contained in the the Solar System giant planets could provide the required depletion \citep{Kunitomo2018}, as long as their volatile content is a factor 5 to 10 lower than predicted by condensation sequences \citep{Lodders2003}. This need to remove volatiles from the giant planets for them to explain the condensation temperature trend has often been used to discard this hypothesis \citep[e.g.][]{Chambers2010}; however, it is possible that outgassing when planetesimals differentiate might reduce the water content enough, if the bodies were initially large ($\gtrsim 50\,{\rm km}$; \citealt{Lichtenberg2019}).

\begin{figure*}
    \centering
    \includegraphics[width=\textwidth]{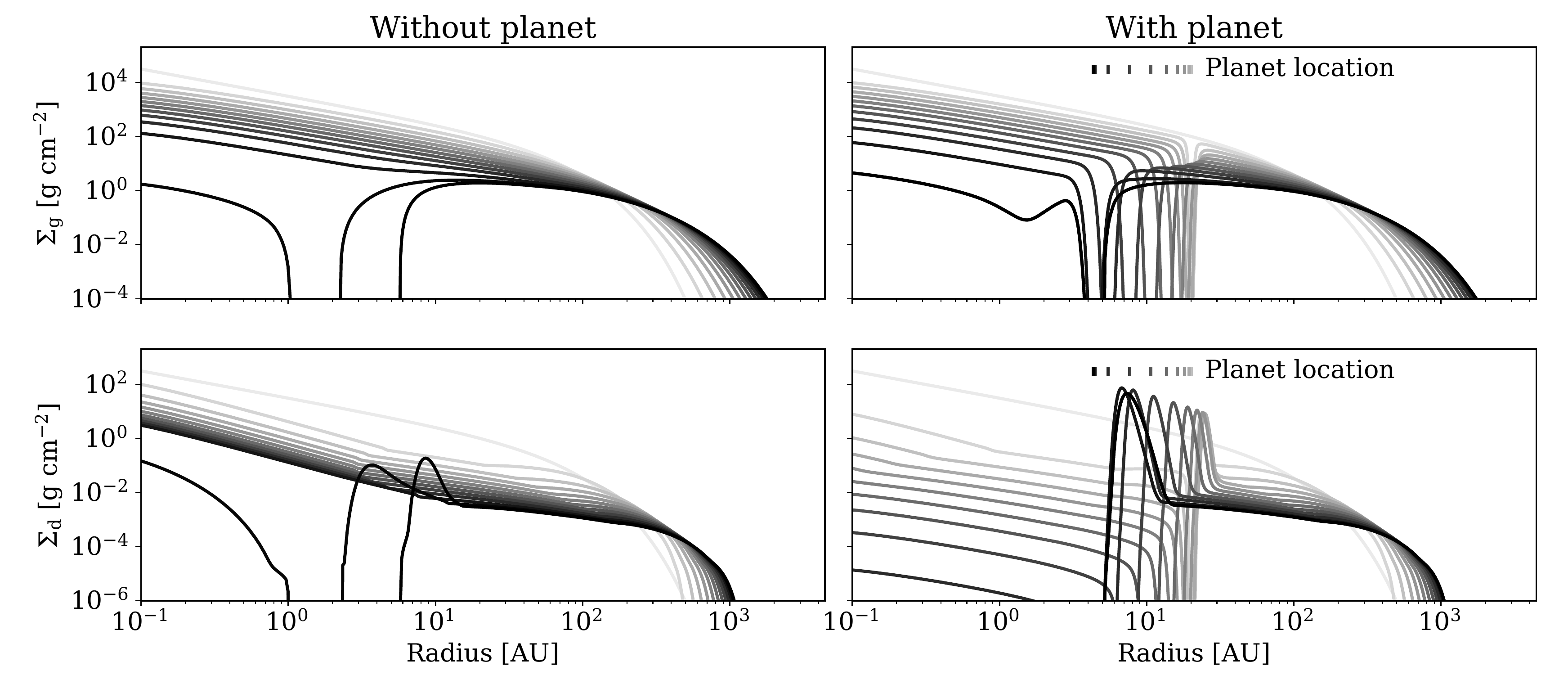}
    \caption{The evolution of the gas and dust surface density in a model with $L_{\rm X} = 4\times 10^{29}\,{\rm erg s}^{-1}$ and $f_{\rm grow} = 10$. A line is plotted every 1~Myr until gas accretion on to the star ceases, with the line shading getting darker as time increases. The left panels show the case without a planet, while the right panels include a planet forming at 0.5~Myr. Note the two order of magnitude difference in the dust mass remaining the disc at the point accretion ceases between the non-planet and planet case. The vertical bars at the top of the two right hand panels show the planet's location.}
    \label{fig:SurfDense_Evol}
\end{figure*}

Finally, we note that the difference in refractory composition has also been interpreted as an enrichment of the twin stars relative to the Sun due to the engulfment of rocky planets. While this may occur in some cases, it seems unlikely to provide the explanation for the bulk of the population since essentially all of the Kepler planet hosting systems would need to have undergone dynamical instability and engulfed a planet. Dynamical instability can already be ruled out for systems with 5 or more planets \citep{Pu2015}, which may account for more than 50 per cent of stars \citep{Zink2019} -- too many for engulfment to be responsible for difference in refractory composition. Therefore, in the next section, we argue instead that the mass of dust trapped in the disc due to gap opening by the planets provides a natural explanation.

\subsection{Depletion due to giant planet formation}

\begin{figure*}
    \centering
    \includegraphics[width=\textwidth]{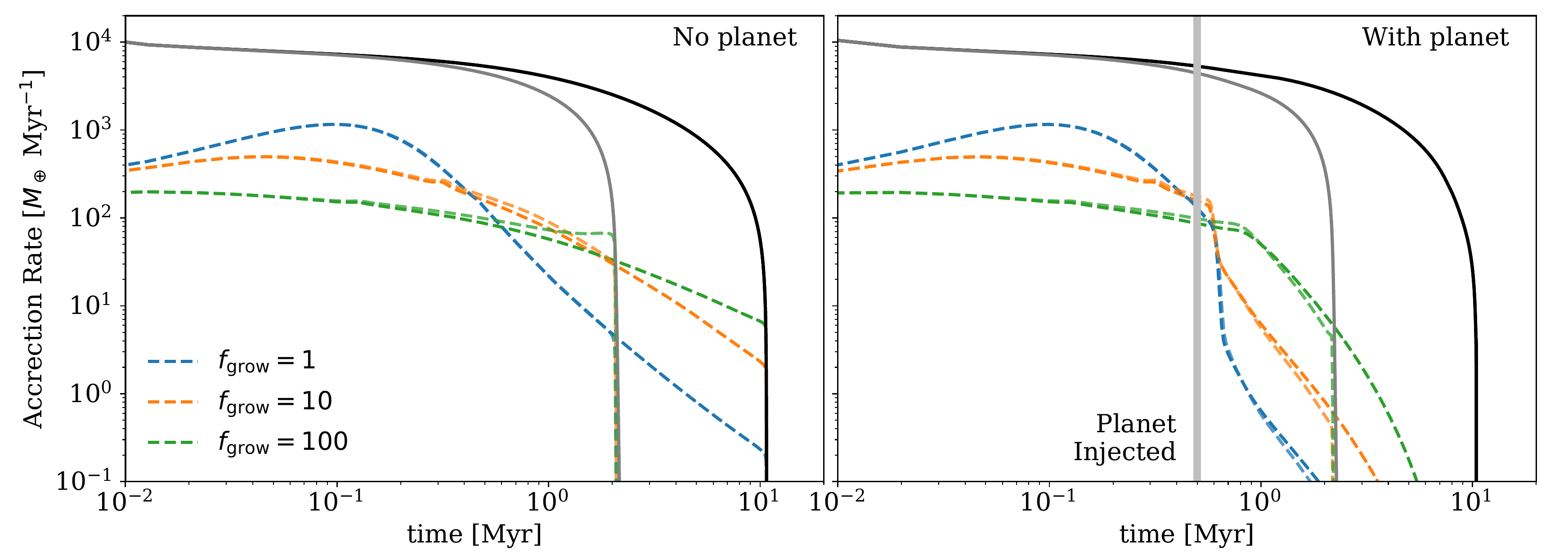}
    \caption{Accretion rate of gas and dust onto the star in models with and without a planet (right and left panels, respectively). The gas accretion rate is shown as solid lines and the dust as dashed lines with lighter and darker lines used for models with high and low $L_{\rm X}$ respectively. Different values of $f_{\rm grow}$ are denoted by different line colours for the dust, (note that the gas accretion rates are independent of $f_{\rm grow}$. The vertical grey line denotes the time when the planet is injected.}
    \label{fig:AccretionRate}
\end{figure*}

\begin{figure*}
    \centering
    \includegraphics[width=\textwidth]{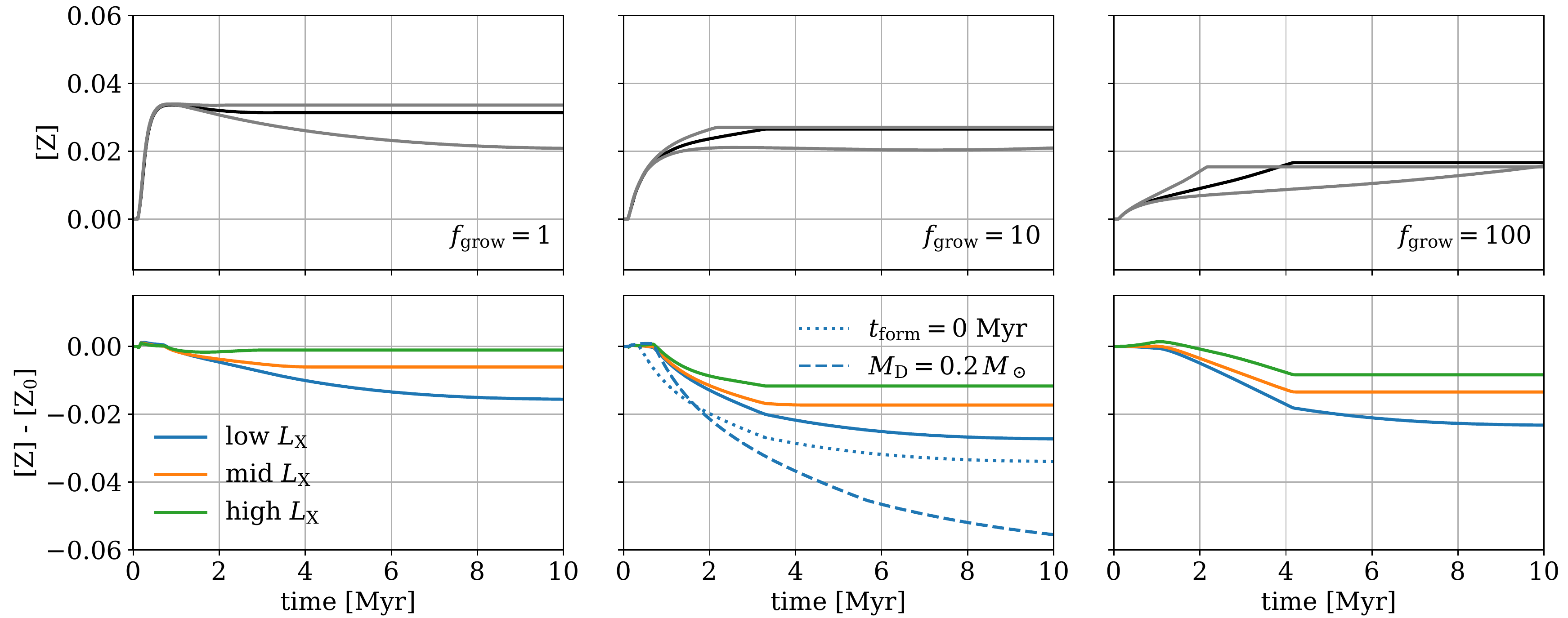}
    \caption{Evolution of the stellar photosphere composition. Top row: the evolution of the photospheric abundances in the reference model without a planet (black lines), representing a `typical' solar twin. For comparison the models with low and high $L_{\rm X}$ are shown as grey lines. Bottom row: the difference in the abundances between the `typical' twin models and the models with a planet. The planet is assumed to form at 0.5~Myr (except in one case).}
    \label{fig:Composition}
\end{figure*}

The evolution of the gas and dust surface density in a typical model is shown in \autoref{fig:SurfDense_Evol} for both the cases without a planet and with planet forming at 0.5~Myr. For this model we assume an X-ray luminosity of $4\times 10^{29}\,{\rm erg s}^{-1}$ and $f_{\rm grow} = 10$. First, we briefly re-cap the evolution of the gas and dust in the absence of a planet, which has been explored in detail by \citet{Alexander2007}, \citet{Owen2011}, \citet{Ercolano2017} and \citet{Owen2019}. The gas evolution is predominantly viscous until the accretion rate becomes comparable to the mass-loss rate due to photoevaporation. At this point photoevaporation opens up a gap in the disc preventing the re-supply of gas. Accretion onto the star terminates once the inner disc has accreted onto the star. Thus, not all the gas initially present accretes onto the star. The dust surface density initially evolves more rapidly than the gas because the grains grow to sizes where they accrete rapidly by radial drift onto the star. Again, once photoevaporation opens a gap in the disc the accretion of dust stops since dust becomes trapped in the pressure maximum outside of the photoevaporative gap.

In simulations that include a planet, the accretion of dust declines rapidly. This occurs again due to the trapping of dust in a pressure maximum, but in this case the trap is associated with the planet rather than photoevaporation. However, since gaps created by planets allow gas to flow through them \citep[e.g.][]{Artymowicz1996}, the accretion of gas continues until it is terminated by photoevaporation ($\approx 10~\rm{Myr}$).

In summary, the faster evolution of dust than gas means that in models without a planet the star tends to accrete a greater fraction of the dust than the gas, leading to an enhancement of the refractory abundance of the star. However, when a giant planet is present the reverse is true because a large fraction of the dust is trapped outside the planet's orbit, which results in a reduction in the refactory abundance of the star.

One of the most uncertain parameters is the dust growth rate (parameterised through $f_{\rm grow}$). The impact of the grain growth rate, is explored in \autoref{fig:AccretionRate}. A faster grain growth rate (low $f_{\rm grow}$) initially leads to a higher accretion rate of dust because grains reach larger sizes, which drift faster. Since radial drift leads to the disc becoming depleted in dust, models with fast growth have lower dust accretion rates after 1~Myr. However, the sensitivity to the dust growth time is fairly small, with the accretion rate varying only by less than a factor of 10 despite $f_{\rm grow}$ differing by a factor 100. The reason behind the smaller range of accretion rates for the dust is that small grains are accreted with along with the gas. This is a consequence of the relative velocity between the dust and gas being small for small grains, leading to the gas and dust accreting at the same velocity. This limits the accretion rate of the dust to be no smaller than the product of the gas accretion rate and the dust-to-gas ratio. \autoref{fig:AccretionRate} shows that $f_{\rm grow}\sim 10$ represents a balance between rapid growth (and hence rapid accretion onto the star) and slow growth where the dust tracks the gas due to its small size.  

\autoref{fig:AccretionRate} also clearly shows the impact of a planet: the accretion rate of dust begins to decline soon after the planet forms, with the decline fastest when the grains are largest. From \autoref{fig:AccretionRate}, we also see that the accretion of dust is primarily controlled by $f_{\rm grow}$ and the presence or absence of a planet. The effect of photoevaporation on the dust is primarily to shut off the accretion of dust once a photoevaporative gap is opened. Conversely, the gas evolution is sensitive the mass-loss via photoevaporation but is independent of $f_{\rm grow}$ (by assumption) and only weakly sensitive to the presence of a planet.

We now explore the resulting effect on the stellar surface composition. Since we are interested in the difference between the Sun's surface composition and the average of the solar twins (0.04 dex, or 10 per cent \citealt{Melendez2009}), we compared the photospheric composition of models with a planet to a reference model without planets and with an X-ray luminosity given by the median of those observed for young stars as a baseline (i.e. $L_{\rm X} = 1.5 \times 10^{30}$~erg~s$^{-1}$), namely a typical disc evolutionary pathway. 

The differential evolution of dust and gas in the absence of a planet gives rise to a baseline composition that differs from the assumed dust-to-gas ratio of the disc (assumed to be 0.01). However, this composition only weakly depends on the dust growth time scale, $f_{\rm grow}$, (\autoref{fig:Composition}) because for $f_{\rm grow} \lesssim 10$ essentially all of the dust mass accretes onto the star before photoevaporation opens a gap. For $f_{\rm grow} = 100$ more dust remains, leading to a slightly lower composition.  By varying the photoevaporation rate (through varying $L_{\rm X}$), we see that the typical star-to-star variation from differences in disc evolution alone should be less than 0.01 dex.

The depletion of refractory material in models with planets with respect to the median star is shown in the bottom row of \autoref{fig:Composition} (at fixed $f_{\rm grow}$). After 10~Myr the composition does not evolve significantly in all models because accretion has terminated. 

The primary factors affecting the difference in composition between the planet-hosting stars and the reference compositions are the amount of dust remaining in the disc when the planet opens a gap and the length of time after this for which gas continues to accrete onto the star (which is controlled by the photoevaporation rate via $L_{\rm X}$). 

For $f_{\rm grow} = 1$ all of the models produce depletions $<0.02$~dex because most of the dust has accreted onto the star by the time the planet opens up a gap and the convective zone is 100\% of the star. However, the amount of dust remaining in the discs (approx $30\,M_\oplus$) is also lower than inferred dust mass in most large-holed transition discs  \citep{Andrews2012,vanderMarel2018}. This result should not be surprising given the fast evolution of dust in these models compared with observations. 

The amount of depletion caused by introducing a planet increases going from $f_{\rm grow}=1$ to 10, but decreases again going from $f_{\rm grow}=10$ to 100. The initial increase going to  $f_{\rm grow}=10$ is caused because the larger dust mass remaining in the disc when the planet opens the gap. However, while the mass remaining in the disc is again higher at $f_{\rm grow}=100$, the reference model without a planet had a lower initial abundance because of more dust mass remaining in the disc when photoevaporation terminates accretion. Hence there is a sweet-spot with regard to the dust growth time-scale. 

We also investigated the effects of the initial dust-to-gas ratio and disc mass on the observed abundance. While changing the dust-to-gas ratio affects the amount of dust accreted by the star, our assumption that the disc and star initially have the same composition means that the change relative to other stars of the same metallicity is small\footnote{Differences in the dust growth time-scale and the fact that the gas mass accreted is assumed to be the same do lead to small changes, but these are less than    $5\times 10^{-3}$ dex}. Conversely, increasing the disc mass naturally leads to a larger depletion of photospheric abundances because the total mass of dust remaining in the disc is larger while the stars initial abundance has not changed.

Finally, we also explored how the depletion depends on where or when the planet was injected. The depletion is not sensitive to our default assumption of a planet forming at 20~au and migrating to Jupiter's current location; formation closer in results in a slightly higher amount of dust mass trapped in the disc and therefore a stronger depletion, while formation further out leads to the opposite. Other migration paths, such as the inward migration followed by outward migration assumed in the `grand tack' model \citep{Walsh2011}, would not lead to significantly different levels of depletion. The presence of giant multiple planets is not expected to make a significant difference either (e.g. \citealt{Hogbolle2019}). Instead, the main factor is the timing of the giant planet formation due to the decline of dust mass as the disc evolves.

Producing the 0.04 dex depletion observed in the Sun relative solar twins relies on what might seem like ``optimistic'' assumptions. Larger disc masses, e.g. $M_{\rm D} \approx 0.2\,M_\odot$, can easily produce an 0.04 dex depletion together with the presence of a planet. Similarly, a low photoevaporation rate (low $L_{\rm X}$) leads to a larger depletion because of the  accretion of more dust-free gas. The depletion is also enhanced by the decreasing size of the convective zone after 2~Myr. Earlier formation of the planet also helps produce a large depletion as more dust mass is trapped in the disc. 

It should not be surprising that our models require ``optimistic'' conditions to achieve a the observed 0.04 dex depletion, given our default assumptions are conservative. The non-accreting stellar model is fully convective for 2~Myr, with the outer convective zone still containing 40 per cent of the stars mass at 10~Myr. Since we assume an initial disc-to-star mass ratio of 0.1-0.2, the 0.02 to 0.06 dex depletion (5 to 15 per cent) shows that giant planets are actually very efficient at causing a depletion in the stellar composition relative to stars without planets. We further emphasise that the conditions that favour large depletions, i.e. ``optimistic'' assumptions (higher disc mass or longer disc lifetime) are those that are expected to favour giant planet formation.

\section{Discussion}
\label{sec:discuss}
We have explored dust trapping by giant planets as a potential explanation for the Sun's peculiar composition, which shows an 0.04 dex (10 per cent) depletion of refractory species with respect to 80 to 90 per cent of the solar twins \citep{Melendez2009,Bedell2018}. We find that the formation of giant planets can easily produce a 0.02 to 0.04 dex depletion with respect to stars without giant planets, even with the conservative assumptions made here (an initial disc-to-star mass ratio of 10 per cent and an solar evolutionary track that remains fully convective for $>2$~Myr). By comparison, the star-to-star variation in abundance from differences in the disc lifetime (due to variations in the photoevaporation rate) is less than 0.01 dex. 

The depletion arises by a combination of two factors: 1) the stellar abundance is slightly enhanced with respect to the initial disc abundance because more gas than dust remains in the disc when accretion is terminated by photoevaporation when no giant planet forms. 2) The formation of planet prevents dust accreting onto the star, while the accretion of refractory-poor gas continues. At low X-ray luminosity, when the disc lifetime is longest, the reduction in refractory abundance is further enhanced due to decrease in the size of the star's convective zone.

A key question that remains is whether the trapping of dust outside the orbit of giant planets produces the correct depletion in terms of condensation temperature, which we have not addressed here. We instead point to the fact that Herbig Ae stars with large cavities in their millimetre emission show depletion of refractory species but not carbon or oxygen \citep{Kama2015}. This is despite a number of the cavities being large enough that water might expect to be frozen out in the disc. Clearly, this question needs to be addressed from a theoretical point of view; however, a plausible explanation is that the inner edge of the cavities are exposed to direct irradiation from the star leading to much higher temperatures and possibly driving volatile species into the gas phase. Simple estimates suggest that the temperatures in the directly irradiation region could be a factor 5 higher \citep{Chiang2001} than typical mid-plane temperatures, but we will address this question further in future work. Furthermore, the transition disc cavity could be optically thin to UV photons, meaning that a large fraction of the disc's gas and dust in the vicinity of the dust trap could be in the warm molecular gas-phase, further driving volatile species into the gas which could then be photo-discossiated preventing them from condensing into the solid phase. Since the dust-trap will move in with time as the planet migrates, exposing it to higher levels of irradiation we speculate that this could explain the condensation trend. As the planet moves closer and closer to the star more refractory elements could be liberated into the gas phase and then accrete onto the star; however, the shorter remaining disc lifetime means less would accrete onto the star producing a trend with condensation temperature. If and how this process happens requires 3D hydrodynamical modelling of the planet-disc interaction with dust-gas chemistry; something that has not been performed. 

There is good reason to believe in the solar system that the formation of Jupiter formed early, as required in our scenario, and had an impact on the composition of the Sun. \citet{Kruijer2017} used isotope ratios of tungsten and molybdenum to show that the reservoirs of material that formed noncarbonaceous and carbonaceous chondrites were separated within 0.5 to 1~Myr, which they suggest was due to gap opening by Jupiter. \citet{Kruijer2017} also suggest hat Jupiter was still growing at 3 to 4~Myr. Further, suggestions from atmospheric escape modelling of the early terrestrial planets suggest disc lifetimes in excess of $4$~Myr \citep[e.g.][]{Lammer2019} meaning that, at the very least, the solar system's protoplanetary disc was not amongst the shorter lived discs. This is promising, given that the largest refractory abundance depletions are produced for disc lifetimes of around 5~Myr or longer. 

The most direct test of whether gap opening by giant planets could be responsible for the depletion would be to see whether the twin stars showing similar levels of depletion to the Sun also host planets. However, the possibility that the planets causing the signature maybe several au from the star makes such a measurement challenging. Current evidence along these lines is not very strong \citep[see][]{Bedell2018}, but if anything maybe in disagreement with our hypothesis because the two stars with candidate planets do not show the depletion in refractory abundances.

Another test can be made using planet-hosting binary twins. Here the results are also mixed, with a number of different interpretations. E.g., the co-moving pair of solar-type stars HD 240429 and HD 240430 show a 0.2 dex difference in abundance that is may be due to engulfment of planetary material \citep{Oh2018}, something that maybe more likely for binary stars due to perturbations to planetary orbits. \citet{Teske2016} show that the stars in the binary HD 133131AB, in which both stars host giant planets, differ in composition at 0.03 dex level, which they suggest could be caused by accounting for the differences in solid mass locked up in the planets. Finally, \citet{Ramirez2011} reported that the giant planet-hosting secondary in the 16 Cyg binary was depleted in metals, possibly due to the giant planet, but did not find evidence for the condensation temperature trend.

Nevertheless, our models show that giant planet formation should leave a signature in the stellar composition, assuming that the planets are able to trap dust outside of their orbits. Even if other processes are acting, the mechanism considered here should contribute to abundance variations at a comparable level.

\section{Conclusions}
\label{sec:conclusion}

Using evolutionary models for the protoplanetary disc that surrounded the young Sun, we have investigated several scenarios for the origin of the solar abundance anomaly, wherein the refractory (rock-forming) elements in the Sun's photosphere are depleted relative to nearby sun-like stars. Since the dust can drift relative to the gas in protoplanetary discs, the dust accretion rate does not trace the gas accretion rate, meaning that the accretion of dust (refractory material) and gas (volatiles) can cause the Sun's refractory to volatile elemental abundances to diverge.  We consider two general disc evolution scenarios, one in which a giant planet formation occurs in the outer disc and another in which giant planet formation does not happen. We explain the solar abundance anomaly in a scenario where a forming giant planet opens a gap in the gas disc, creating a pressure maxima exterior to it is orbit. This pressure maxima traps the dust in the protoplanetary disc, preventing it accreting onto the star while continuing to let gas accrete on to the star. Our main conclusions are summarised below:
\begin{enumerate}
    \item When a star does not form a giant planet at large separations the dust rapidly grows and drifts towards the star, resulting in an enhanced accretion of dust relative to gas at the $\sim 0.02-0.03$~dex level. 
    \item We demonstrate that the sequestering refractories in rocky planets is not the origin of the refractory depletion. We show the fact that exoplanet surveys (such as {\it Kepler}) have shown the majority of Sun-like stars possess close-in planetary systems that are common, and on average more massive than our Solar System's, implies that the Sun's depletion cannot be caused by the sequestration of refractory elements in the terrestrial planets.  
    \item When a star does form a giant planet at large separations dust is trapped in a pressure trap exterior to the planet's orbit, and the star accretes more gas than dust from the protoplanetary disc. If the planet forms early ($\lesssim 1$~Myr) while the disc is still massive, $\gtrsim 100$~M$_\oplus$ of dust can be trapped and prevented from accreting onto the star.  
    \item We argue the Sun's (and other similarly refractory poor Solar twins) giant planets trapped dust outside their orbits while still forming, preventing the refractory material accreting onto the star. Whereas the majority of the Solar twins did not form a giant planet, hence they accreted the majority of dust in their disc. This mechanism results in refractory poor Stellar abundances at the level of 0.02-0.06~dex for stars that form giant planets. 
    \item Higher refractory depletions are favoured for more massive, and longer-lived discs, both of which favour giant planet formation. 
\end{enumerate}
While our model of trapping dust outside a forming giant planet is successful in reproducing the level of required refractory depletion, it is not sophisticated enough to model the condensation trends of individual elements. Such detailed comparison motivates the need for the development of more sophisticated models in two different areas. Firstly, detailed chemical evolution models for the evolution and transport of refractory and volatile elements in an evolving protoplanetary disc along the lines of those by \citet{Booth2017,Booth2019} rather than just treating ``gas'' and ``dust'' as we have done in this work. Secondly, there is almost no understanding of the role of planet-disc interactions in controlling how different species may enter the gas and flow across the planetary gap, or be accreted by the planet's themselves. Finally, as indicated by observations of transition discs around Herbig Ae/Be stars, the smaller (or non-existent) convective zone sizes around stars more massive than the Sun motivate extending this work to more massive stars where the effects are likely to be more pronounced.     

\section*{Acknowledgements}
We acknowledge the reviewer for a helpful review that improvement the manuscript.
RAB acknowledges support from the STFC consolidated grant ST/S000623/1. JEO is supported by a Royal Society University Research Fellowship. This work has also been supported by the European Union's Horizon 2020 research and innovation programme under the Marie Sklodowska-Curie grant agreement No 823823 (DUSTBUSTERS). We thank the organizers of the  Star-Planet 2019 conference at Ringberg, where the initial ideas for this paper were formed. We also thank A. Korn, O. Shorttle and M. Kama for interesting discussions.




\bibliographystyle{mnras}
\bibliography{twins} 



\appendix


\bsp	
\label{lastpage}
\end{document}